\documentclass[runningheads]{llncs}
\usepackage[utf8]{inputenc}
\usepackage{array}
\usepackage{verbatim}
\usepackage{scalerel}
\usepackage{mathtools}
\usepackage{graphicx}
\usepackage{algorithm}% http://ctan.org/pkg/algorithms
\usepackage{algpseudocode}% http://ctan.org/pkg/algorithmicx
\makeatletter

\usepackage{pgfplots}
\usepackage{subcaption}
\usepackage{adjustbox}
\usepackage{url}

\usepgfplotslibrary{groupplots,dateplot}
\usetikzlibrary{
    patterns,
    chains,
    backgrounds,
    calc,
    shadings,
    shapes.arrows,
    arrows,
    shapes.symbols,
    shadows,
    positioning,
    decorations.markings,
    backgrounds,
    arrows.meta,
    external
}
\usepackage{array}
\usepackage{algorithmicx,algpseudocode}
\algblock{ParFor}{EndParFor}
\algnewcommand\algorithmicparfor{\textbf{parfor}}
\algnewcommand\algorithmicpardo{\textbf{do}}
\algnewcommand\algorithmicendparfor{\textbf{end}}
\algrenewtext{ParFor}[1]{\algorithmicparfor\ #1\ \algorithmicpardo}
\algrenewtext{EndParFor}{\algorithmicendparfor}

\pgfplotsset{compat=newest}

\newif\iffinal

\iffinal
  \newcommand{\maxx}[1]{}
  \newcommand{\ryan}[1]{}
  \newcommand{\todo}[1]{}
\else
  \newcommand{\maxx}[1]{{\textcolor{red}{ Max: #1 }}}
  \newcommand{\ryan}[1]{{\textcolor{magenta}{ Ryan: #1 }}}
  \newcommand{\todo}[1]{{\textcolor{blue}{ TODO: #1 }}}
\fi

\makeatother

\begin{document}
\title{Ultrafast Focus Detection for Automated Microscopy}
% \author{Maksim Levental}
% \affiliation{\institution{University of Chicago}}
% \author{Ryan Chard}
% \affiliation{\institution{Argonne National Laboratory}}
% \author{Gregg A. Wildenberg}
% \affiliation{\institution{University of Chicago}}

\author{Maksim Levental \inst{1} \and
    Ryan Chard\inst{2} \and
  Kyle Chard\inst{1,2} \and
  Ian Foster\inst{1,2}
  Gregg Wildenberg \inst{2}
}
\authorrunning{M. Levental et al.}
% First names are abbreviated in the running head.
% If there are more than two authors, 'et al.' is used.
%
\institute{University of Chicago, Chicago IL, USA
  \and
  Argonne National Lab, Lemont IL, USA\\
   %\email{@uchicago.edu}
}

\maketitle

\begin{abstract}
  % Scientific instrument have seen a 
%   There has been a dramatic increase in the volumes and velocities of data generated by modern scientific instruments. % deployed in every-day laboratories.
%   Scanning electron microscopy (SEM) is one such example where technological advancements have led to scientists being overwhelmed with critical data for montaging, alignment, and image segmentation---key
%   practices for many scientific domains, including, for example, neuroscience.
  Technological advancements in modern scientific instruments, such as 
  scanning electron microscopes (SEMs), have significantly increased data acquisition rates and image resolutions enabling new questions to be explored; however, the resulting data volumes and velocities, combined with automated experiments, are quickly overwhelming scientists as there remain crucial steps that require human intervention, for example reviewing image focus.
 %   , where SEM are used to derive the anatomical
%   relationships of the brain.
  % These instruments now necessitate equally advanced computing 
  % resources and techniques to realize their full potential.
  We present a fast out-of-focus detection algorithm for electron microscopy images collected serially and demonstrate that it can be used to provide near-real-time quality control for neuroscience workflows.
  Our technique, \textit{Multi-scale Histologic Feature Detection}, adapts classical computer vision techniques and is based on detecting various fine-grained histologic features.
  We exploit the inherent parallelism in the technique to employ GPU primitives in order to accelerate characterization.
  We show that our method can detect of out-of-focus conditions within just 20ms.
  To make these capabilities generally available, we deploy our feature detector as an on-demand service and 
  %at the Argonne Leadership Computing Facility. 
  %We 
  show that it can be used to determine the degree of focus in approximately 230ms, enabling near-real-time use.
  % We discuss extensions that enable scaling out to support multi-beam
  % microscopes and integration with existing focus systems for purposes
  % of implementing auto-focus.

\end{abstract}
% \settopmatter{printfolios=true}

\section{Introduction}

\label{sec:intro}

% \begin{comment}
% Advancements in the automation of serial scanning electron microscopy
% (SEM) impose a regime where thousands, if not tens of thousands, of
% images can now be automatically collected by researchers. This puts
% greater demand on conventional auto-focus algorithms for ensuring
% each image is in focus, as an alternative to the user manually evaluating
% each image by eye. Without such algorithms, critical bottlenecks are
% created where the user is forced to reacquire individual, deficient
% (out-of-focus), images and manually reinsert them into the sequence
% of thousands of other images already acquired. This is an onerous
% task which requires taking into account alignment and boundary overlap.
% Furthermore, failure to quickly identify and reacquire deficient images
% negatively impacts the accuracy of downstream, post-processing; for
% example 2D montaging, 3D alignment, or automatic segmentation pipelines.
% While many microscopes have builtin auto-focus algorithms, these often
% fail to achieve acceptable accuracy due to intrinsic mediating factors
% (e.g. stage drift) and extrinsic mediating factors (e.g. sample artifacts,
% non-uniformity in the sample).
% \end{comment}

A fundamental goal of neuroscience is to map the anatomical relationships
of the brain, an approach broadly called \textit{connectomics}. 
Electron microscopy, an imaging method traditionally limited to small single 2D images, provides sufficient resolution to directly visualize the connections, or synapses, between neurons. 
Recently, automated serial electron microscopy (SEM) techniques have been developed where thousands, if not tens of thousands, of individual images are automatically acquired in series and then registered (i.e., aligned) to produce a volumetric dataset.
Such datasets allow neuroscientists to follow the tortuous path neurons take through the brain to connect with each other (hence the name connectomics). 
However, many of the steps that comprise the collection of such datasets for connectomics require manual inspection, causing significant slowdowns in the rate at which datasets can be acquired.
Such bottlenecks significantly impact the size of the datasets that can be reasonably acquired and studied. 
Furthermore, advances in electron microscopes have increased the rate that datasets can be acquired; for example, $\sim$ 10 Tbs/24hr \cite{zeiss:multisem550}, which, when used to map an entire, mouse brain will result in approximately 1 exabyte of data. 
% Thus, there is a natural need for automation in the collection and processing of SEM data.

Auto-focus technology is a critical component of many imaging systems; from consumer cameras (for purposes of convenience) to industrial inspection tools to scientific instrumentation~\cite{1545017}.
Such technology is typically either \textit{active} or \textit{passive}; active methods exploit some auxiliary device or mechanism to measure the distance of the optics from the scene, while passive methods analyze the definition or sharpness of an image by virtue of a proxy measure called a \textit{criterion function}. 
Many electron microscopes incorporate auto-focus systems that attempt to focus the microscope before image acquisition. 
Despite such functionality, out-of-focus (OOF) images still occur at high rates (between 1\% and 10\%), depending on the quality of the tissue sections being imaged. 
For instance, it is common to experience occasional staining artifacts, and tears or compression artifacts (i.e., section wrinkles) during ultra-thin serial sectioning. 
These imperfections can cause auto-focus systems to fail if the microscope centers on them.
This results in the system failing to find the correct focal plane, thus necessitating post-acquisition evaluation. 
These OOF error modes prevent effective automation, since a prerequisite of many downstream transformations is that the images collected all have high degree-of-focus (DOF). 
Without properly focused images, all downstream computational steps (e.g., 2D tile montaging, 3D alignment, automatic segmentation) will fail.

The DOF of images acquired by an electron microscope is also of critical importance with respect to automation.
While seemingly a small step in a potential automation pipeline, focus detection is nevertheless an extremely critical step.
In general, imaging tissue sections requires loading and unloading sets of $\sim$ 100-200 sections at a time.
Failure to detect a single OOF image in situ causes significant delays because the affected sample sets need to be reloaded, desired field of view must be reconfigured, and reacquired images need to be realigned into the image stack.
All such remediation steps are time and labor intensive, and effectively stops any downstream automation until the problem is remedied. 
Under ideal conditions, it is estimated that fixing a single image would take several hours of manual intervention, which increases if multiple images in distinct parts of the series have to be manually reacquired and aligned.

In this work we focus on ensuring images acquired by the electron microscope have high DOF, in order to further progress towards to goal of end-to-end automation. 
To this end, we propose a new technique, \textit{Multi-scale Histologic Feature Detection} (MHFD), that involves a second pass over the collected image, after it has been acquired, using a computer vision system to detect a failure to successfully achieve high DOF. 
Our technique relies on employing feature detection~\cite{Lindeberg2004FeatureDW} as a criterion function, in accordance with the hypothesis that the quantity of features detected is positively correlated with DOF. 
Using this insight, we develop a feature detector based on scale-space representations of images (see Section~\ref{subsec:scalespace}) but optimized for latency. 
The design and implementation of our feature detector prioritizes parallelization, specifically in order to target GPU deployments.

Our solution achieves low latency detection of the OOF condition with high accuracy (see Section~\ref{sec:Evaluation}).
To provide access to these capabilities, we have deployed them as a service that can be consumed on-demand and integrated in automated workflows.
The service leverages Argonne National Laboratory's Leadership Computing Facility to provide access to A100 GPUs to rapidly analyze images as they are captured.
% Further, these resources can be employed in
% automated pipelines
% to reliably and securely perform quality control on the data as they are created.
% Our pipeline rapidly alerts users to low quality images and enables them to 
This allows users to detect low quality images and correct their collection while the sample is still in the microscope, effectively eliminating costly delays in reloading, aligning, and imaging the sample.
An important caveat in our work: we explicitly aim to augment existing microscopy equipment without the need for costly and complex retrofitting. 
This precludes mere improvements to existing auto-focus systems as they are, in essence, proprietary black boxes from the perspective of the end user of an electron microscope.

The rest of this article is organized as follows: Section~\ref{sec:background} reviews background information on connectomics and scale-space feature detectors. Section~\ref{sec:mhfd} describes our focus detection method, in particular optimizations made in order to achieve
near-real-time performance. 
Section~\ref{sec:service} describes how we deliver MHFD as a service. 
Section~\ref{sec:Evaluation} presents
evaluation results. % Section~\ref{sec:Discussion} discuss those results,
Section~\ref{sec:related} discusses related work. Finally, we conclude
in Section~\ref{sec:conclusion}. %how our work is distinct therefrom.

\section{Background}\label{sec:background}

We briefly review a common connectomics workflow and then describe scale-space representations.

\subsection{Connectomics}\label{subsec:Connectomics}

Connectomics is defined as the study of comprehensive maps of connections within an organism's nervous system (called \emph{connectomes}).
The data acquisition pipeline for connectomics consists of the following steps:
\begin{enumerate}
    \item A piece of nervous system (e.g., brain), ranging from $\sim$ 1mm$^3$ to 1 cm$^3$ is stained with heavy metals (e.g., osmium tetroxide, uranyl acetate, lead) in order to provide contrast in resulting images~\cite{doi.org/10.1038/ncomms8923};
    \item After staining, the section is dehydrated and embedded in a plastic resin to stabilize the tissue for serial sectioning, which is performed with an Automated Serial Sections to Tape (ATUM) device~\cite{kasthuri2015saturated} (where ultrathin sections are automatically sectioned and collected on polyimide tape);
    \item The sections are mounted to a silicon wafer, with each wafer containing 200-300 sections;
    \item The wafer is loaded into a SEM, where the user marks a region of interest (ROI) within the sections for the microscope to image;
    \item The SEM initiates a protocol to automatically image the ROI over all the sections at a desired resolution;
    \item For each section, the SEM attempts to auto-focus before imaging by sampling different focal planes over a set range of focal depths
\end{enumerate}
The series of collected images are then algorithmically aligned to each other to produce a 3D volumetric image stack where biological features are segmented either manually or by automatic segmentation techniques. 

Since the imaging and post-acquisition process (e.g., retakes of blurry images, 3D alignment, segmentation) is slow, connectomics is practically constrained to small volumes ($\sim$ 100 µm$^3$), but technologies are rapidly advancing, with near future goals of mapping an entire mouse brain~\cite{mindofmouse}. 
Even with 100 µm$^3$ volumes, the scope of the biological problem is large. 
For instance, a single mouse neuron is estimated to receive $\sim$ 5000-7000 connections~\cite{wildenberg2021primate} and the cell density of the mouse cortex is $\sim1.5 \times 10^5$ cells/mm$^3$~\cite{herculano}. 
A 100 µm$^3$ volume will therefore contain $\sim$ 150 neurons receiving 7.5$^5$ synapses, all of which neuroscientists seek to automatically segment and study. 
For an entire mouse brain, there are $\sim7 \times 10^6$ neurons and $3.5 \times 10^10$ connections. 
Ensuring that automatic segmentation algorithms accurately segment neurons depends on having the highest possible quality images and any error in image quality is very likely to produce segmentation errors that propagate in a non-linear fashion. 
For instance, if the connection between two neurons is improperly assigned, the other neurons that those pair of neurons connect to will also be improperly connected, and so on. 
Not only is the biological scope of the problem large, but datasets are also large, both in terms of the number of images and data size. 
Again, using the range of 100 µm$^3$ to 1 cm$^3$ datasets, these volumes will equate to 2,500 to 250,000 sections and $\sim$ 0.7 terabytes or 1 exabyte of data, respectively. 
Thus, the scope of the data both in terms of the biological goal and data management demands automation in the connectomic pipeline, in order to minimize errors and the need for manual OOF detection and correction.

\subsection{Scale-space representations\label{subsec:scalespace}}

We base our multi-scale histologic feature detection technique on classical scale-space representations of signals and images. 
We give a brief overview (see \cite{Lindeberg2004FeatureDW} for a more comprehensive review).
The fundamental principle of scale-space feature detection is that
natural images possess structural features at multiple scales and features at a particular scale are isolated from features at other scales. 
Thus, any image $I\left(x,y\right)$ can be transformed into a scale-space representation $L\left(x,y,t\right)$, where $L\left(x',y',t'\right)$ represents the pixel intensity at pixel coordinates $\left(x',y'\right)$ and \emph{scale} $t'$.
How to construct the representation of the image at each scale is discussed below. 
More importantly, such a representation lends itself readily to scale sensitive feature detection, owing to the fact that features at a particular scale are decoupled from features at other scales, thereby eliminating confounding detections. 
Examples of structural features that can be detected and characterized using scale-space representations include edges, corners, ridges, and so called blobs (roughly circular regions of uniform intensity).

A scale-space representation at a particular scale is constructed by convolution of the image with a filter that satisfies the following constraints: non-enhancement of local extrema, scale invariance, and rotational invariance. Other relevant constraints are discussed in \cite{duits2004axioms}. 
One such filter is the symmetric, mean zero, two dimensional, Gaussian filter~\cite{koenderink1984structure}:
\[
  G\left(x,y,\sigma\right)\coloneqq\frac{1}{2\pi\sigma^{2}}e^{-\frac{x^{2}+y^{2}}{2\sigma^{2}}}
\]
The scale-space representation $L(x,y,t)$ of an image $I(x,y)$ is defined to be the convolution of that image with a mean zero Gaussian filter:
\[
  L\left(x,y,t\right)\coloneqq G\left(x,y,t\right)*I\left(x,y\right)
\]
where $t$ determines the scale. 
$L(x,y,t)$ has the interpretation that image structures of scale smaller than $\sqrt{t^{2}}=t$ have been removed due to blurring. 
This is due to the fact that the variance of the Gaussian filter is $t^{2}$ and features of this scale are therefore ``beneath the noise floor'' of the filter or, in effect, suppressed by filtering procedure. 
A corollary is that features with approximate length scale $t$ will have maximal response upon being filtered by $G(x,y,t)$. 
That is to say, for a $t$ scale feature at pixel coordinates $\left(x,y\right)$ and for scales $t'<t<t''$ we have
\[
  L\left(x,y,t'\right)<L\left(x,y,t\right)<L\left(x,y,t''\right)
\]
This is due to the fact that for scales $t'<t$, small scale features will dominate the response and for $t<t''$, as already mentioned, the feature will have been suppressed.

Note that the aforementioned presumes having identified the pixel coordinates $\left(x,y\right)$ as the locus of the feature. 
Hence, in order to detect features across both scale and space dimensions, maximal responses in spatial dimensions $\left(x,y\right)$ need to also be characterized.
For such characterization one generally employs standard calculus, in order to identify critical points of the second derivatives of $L\left(x,y,t\right)$.
Hence, we can construct scale-sensitive feature detectors by considering critical points of linear and non-linear combinations of spatial derivatives $\partial_{x},\partial_{y}$ and derivatives in scale $\partial_{t}$.
For example the scale derivative of the Laplacian
\begin{equation}
  \partial_{t}\nabla^{2}L\coloneqq\partial_{t}\left(\partial_{x}^{2}+\partial_{y}^{2}\right)L\label{eqn:blobdetector}
\end{equation}
effectively detects regions of uniform pixel intensity (i.e., blobs).

Equation~\eqref{eqn:blobdetector} permits a discretization called \textit{Difference of Gaussians} (DoG)~\cite{marr1980theory}:
\[
  t^{2}\nabla^{2}L\approx t\times\left(L\left(x,y,t+\delta t\right)-L\left(x,y,t\right)\right)
\]
Therefore, we define the following parameters:
$n$, which determines the granularity of the scales detected; $\min_{t}$, the minimum scale detected; $\max_{t}$, the maximum scale detected;  $\delta t\coloneqq\left(\max_{t}-\min_{t}\right)/n$; $t_{i}\coloneqq\min_{t}+\left(i-1\right)\times\delta t$, the discrete
        scales detected.
% \begin{itemize}
%   \item $n$, which determines the granularity of the scales detected
%   \item $\min_{t}$, the minimum scale detected
%   \item $\max_{t}$, the maximum scale detected
%   \item $\delta t\coloneqq\left(\max_{t}-\min_{t}\right)/n$
%   \item $t_{i}\coloneqq\min_{t}+\left(i-1\right)\times\delta t$, the discrete
%         scales detected
% \end{itemize}
% With the above parameters defined, 
We then define the discretized DoG filter:
\begin{equation}
  \operatorname{DoG}\left(x,y,i\right)\coloneqq t_{i}\times\left(L\left(x,y,t_{i+1}\right)-L\left(x,y,t_{i}\right)\right)\label{eqn:dog}
\end{equation}
This produces a sequence $\left\{ \operatorname{DoG}\left(x,y,i\right)\mid i=1,\dots,n\right\}$ of filtered and scaled images (called a Gaussian pyramid~\cite{derpanis2005gaussian}).
Note that there are alternative conventions for how each difference in the definition of $\operatorname{DoG}\left(x,y,i\right)$ should be scaled (including partitioning into so called \textit{octaves}~\cite{1095851}); we observe that linear scaling is sufficient, in terms of accuracy and complexity, for the purposes of detecting OOF conditions.

\section{Multi-scale Histologic Feature Detection}\label{sec:mhfd}

\begin{figure}
  \centering
  \begin{subfigure}[t]{0.46\textwidth}
    \includegraphics[width=1\linewidth]{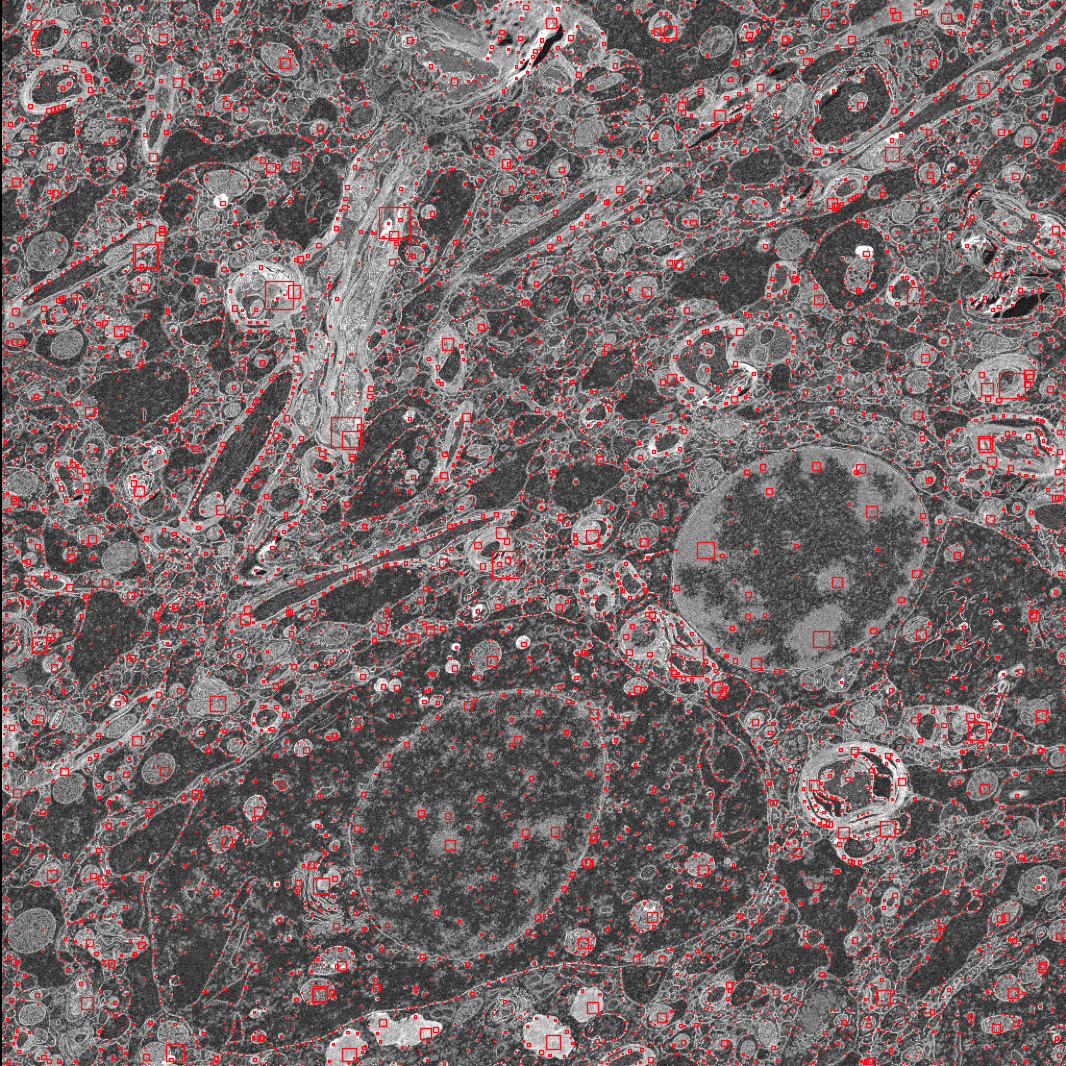}
    \caption{Histologic features of an in-focus section.}
    \label{subfig:infocus}
  \end{subfigure}
  \hfill
  \begin{subfigure}[t]{0.46\textwidth}
    \includegraphics[width=1\linewidth]{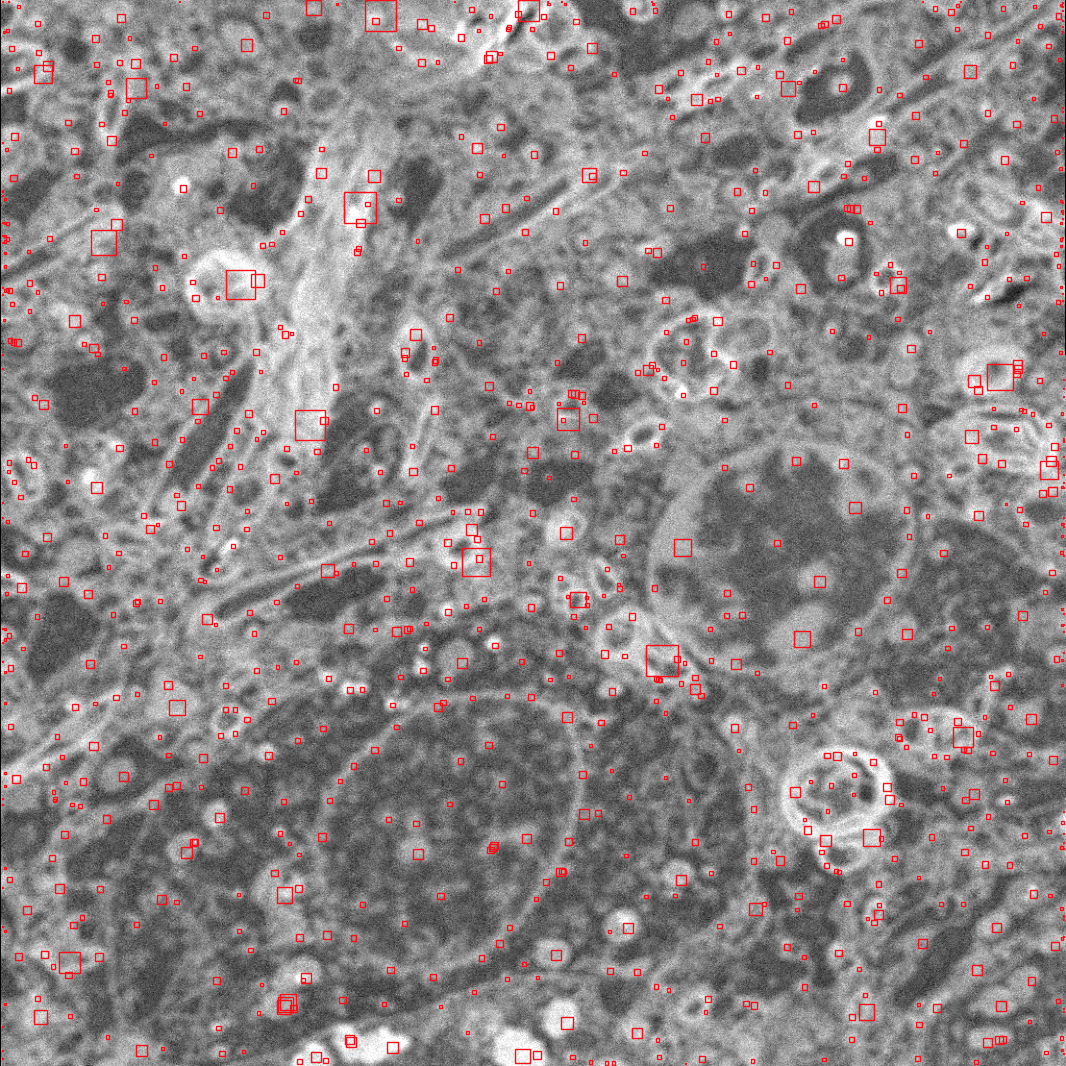}
    \caption{Histologic features of an out-of-focus section.}
    \label{subfig:outoffocus}
  \end{subfigure}
  \caption{Comparison of sections with histologic feature recognition as a function
    of focal depth.}
  \label{fig:histfeatsimages}
\end{figure}

We propose to use histologic feature detection at multiple scales as a criterion function, reasoning that the absolute quantity of features detected at multiple scales is positively correlated with DOF (see Figure~\ref{fig:histfeatsimages}). 
For our particular use case, this is tantamount to detecting histologic structures ranging from cell walls to whole organelles. 
The key insight is that the ability to resolve structure across the range of feature scales is highly correlated with a high-definition image. 
To this end, we develop a feature detector based on Equation~\eqref{eqn:blobdetector} but optimized for latency (rather than accuracy).

\begin{figure}
  \centering
  \begin{subfigure}[t]{0.47\textwidth}
    \includegraphics[width=1\linewidth]{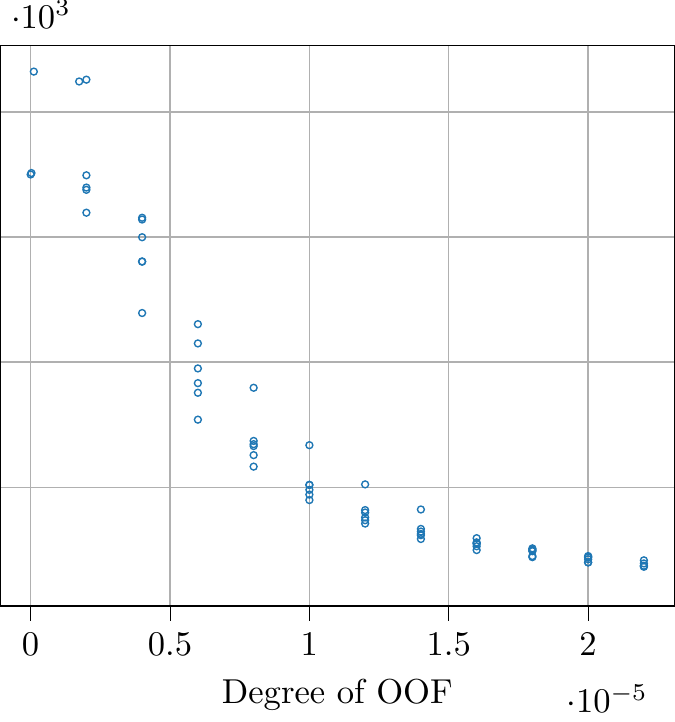}
    \caption{Number of histologic features as a function of absolute deviation
      from focused ($\lvert f-f'\rvert$ where $f'$ is the correct focal
      depth).}
    \label{subfig:degreeoofcurve} \end{subfigure}
  \hfill
  \begin{subfigure}[t]{0.47\textwidth}
    \includegraphics[width=1\linewidth]{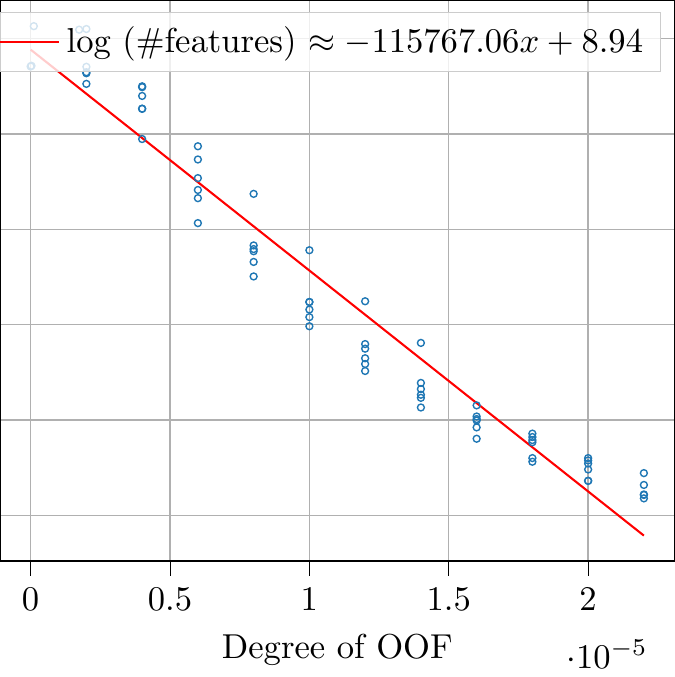}
    \caption{Log plot and line fit with $r=-0.9754$.}
    \label{subfig:degreeooffit} \end{subfigure} \caption{Comparison of histologic feature recognition as a function of focal
    depth.}
  \label{fig:histfeats}
\end{figure}

In order to verify our hypothesis, that detecting features across a range of scales is correlated with DOF, we compare the number of histologic features detected as a function of absolute deviation from in-focus ($\lvert f-f'\rvert$ where $f'$ is the correct focal depth) for a series of sections with known focal depth (see Figure~\ref{subfig:degreeoofcurve}). 
We observe a strong log-linear relationship (see Figure~\ref{subfig:degreeooffit}). 
Fitting a log-linear relationship produces a line with $r=-0.9754$, confirming our hypothesis that quantity of histologic features detected is a good proxy measure for DOF. 
Note that the log-linear relationship corresponds to a roughly quadratic decrease in the number of histologic features detected. 
This is to be expected since, intuitively, a twice improved DOF of a two dimensional image yields improved detection along both spatial dimensions and thus a four times increased quantity of histologic features detected.

Recall, we aim to achieve near-real-time quality control of SEM images to facilitate error detection and correction while data are being collected. 
We therefore require low-latency multi-scale histologic feature detection. 
Here we present the design and implementation of our system that leverages GPUs to rapidly classify images by determining their degree of focus. 
Our starting point is Equation~\eqref{eqn:dog} for possible optimizations.
% We then present optimizations to improve inference latency. Finally, we describe how we have made our system generally available using a function-as-a-service platform.
Computing maxima of $\operatorname{DoG}\left(x,y,i\right)$ in the scale dimension (equivalently critical points of Equation~\eqref{eqn:blobdetector}) necessarily entails computing maxima in a small pixel neighborhood at every scale.
We first make the heuristic assumption that, in each pixel neighborhood that corresponds to a feature, there is a single unique and maximal response at some scale $t$. 
This response corresponds to the scale at which the variance of the Gaussian filter $G$ most closely corresponds to the scale of the feature (see Section~\ref{subsec:scalespace}).
We therefore search for \emph{local maxima} in spatial dimensions $x,y$ but \emph{global maxima} in the scale dimension:
\begin{equation}
  C := \left\{ \left(\hat{x}_{j},\hat{y}_{j},\hat{i}_{j}\right)\right\} \coloneqq\operatorname*{\mathtt{argmaxlocal}}_{x,y}\operatorname*{\mathtt{argmax}}_{i}\operatorname{DoG}\left(x,y,i\right)\label{eqn:argmax}
\end{equation}
where the subscript $j$ indexes over the features detected. 
Once all such maxima are identified it suffices to compute and report the cardinality, $\left|C\right|$, as the criterion function value.

% We now discuss practical optimizations. 
It is readily apparent that our histologic feature detector is parallelizable: for each scale $t_{i}$ we can compute $L\left(x,y,t_{i}\right)$ independently of all other $L\left(x,y,t_{j}\right)$ (for $j\neq i$).
A further parallelization is possible for the $\operatorname*{\mathtt{argmax}}$ operation, since the maxima are computed independently across distinct neighborhoods of pixels. 
In order to maximally exploit this, we first perform the inner $\operatorname*{\mathtt{argmax}}$ in Equation~\eqref{eqn:argmax} on a block of columns of $\left\{ \operatorname{DoG}\left(x,y,i\right)\right\}$ in parallel, thereby effectively reducing the Gaussian pyramid to a single image. 
Note that when GPU memory is sufficient we can compute the $\operatorname*{\mathtt{argmax}}$ across all columns simultaneously (and otherwise within a constant number of steps). 
We then perform the outer $\operatorname*{\mathtt{argmaxlocal}}_{x,y}$ on disjoint pixel neighborhoods of the flattened image in parallel as well.

Note that the implementation of the inner $\operatorname*{\mathtt{argmax}}$ is ``free'', since the $\operatorname*{\mathtt{argmax}}$ primitive is implemented in exactly this way in most GPGPU libraries \cite{CUB}, and thus our substitution of $\operatorname*{\mathtt{argmax}}_{i}$ for $\operatorname*{\mathtt{argmaxlocal}}_{i}$ yields a moderate latency improvement. 
The outer $\operatorname*{\mathtt{argmaxlocal}}$ is implemented using a comparison against $\operatorname{\mathtt{maxpool\_2d}}(n,n)$ (with $n=3$) (see \cite{9307788} for details on this technique).
Employing $\operatorname{\mathtt{maxpool\_2d}}$ in this way has the added benefit of effectively performing non-maximum suppression~\cite{1699659}, since it rejects spurious candidate maxima within a $3\times3$ neighborhood of a true maximum.

Typically one would compute $L(x,y,t_{i})$ in the conventional way (by linearly convolving $G$ and $I$) but prior work has shown that performing the convolution in the Fourier domain is much more efficient~\cite{9307788}; namely
\[
  L\left(x,y,t_{i}\right)=\mathcal{F}^{-1}\big\{\mathcal{F}\left\{ G\left(x,y,t_{i}\right)\right\} \cdot\mathcal{F}\left\{ I\left(x,y\right)\right\} \big\}
\]
where $\mathcal{F}\left\{ \cdot\right\} ,\mathcal{F}^{-1}\left\{ \cdot\right\}$ are the Fourier transform and inverse Fourier transform, respectively.
This approach has the additional advantage that we can make use of highly optimized Fast Fourier Transform (FFT) routines made available by GPGPU libraries.

One remaining detail is histogram stretching of the images. 
Due to the dynamic range (i.e., variable bit depth) of the microscope, we need to normalize the histogram of pixel values.
We implement this normalization by saturating $.175\%$ of the darkest pixels, saturating $.175\%$ of the lightest pixels, and mapping the entire range to $[0,1]$. 
We find this gives us consistently robust results with respect to noise and anomalous features. 
This histogram normalization is also parallelized using GPU primitives.
We present our technique in Algorithm~\eqref{alg:Multi-scale-Histologic-Feature}.

\begin{algorithm}
  \caption{Multi-scale Histologic Feature Detection\label{alg:Multi-scale-Histologic-Feature}}

  \begin{algorithmic}[1]
    \Require{$I\left(x,y\right), n, \min_{t}, \max_{t}, M$}
    \State $I'\left(x,y\right) \coloneqq \texttt{HistorgramStretch}(I\left(x,y\right))$
    \State \texttt{Broadcast}$\left(I'\left(x,y\right), M\right)$
    \ParFor{$m \coloneqq 1,\dots, M$}
    \ParFor{$i\in I_{m}$}
    \State $L\left(x,y,t_{i}\right) \coloneqq \mathcal{F}^{-1}\big\{\mathcal{F}\left\{ G\left(x,y,t_{i}\right)\right\} \cdot\mathcal{F}\left\{ I'\left(x,y\right)\right\} \big\}$
    \EndParFor
    \EndParFor
    \State \texttt{Gather}$\left(L\left(x,y,t_{i}\right), M\right)$
    \ParFor{$i \coloneqq 1,\dots, n+1$}
    \State $\operatorname{DoG}\left(x,y,i\right)\coloneqq t_{i}\times\left(L\left(x,y,t_{i+1}\right)-L\left(x,y,t_{i}\right)\right)$
    \EndParFor
    \State $\left\{ \left(\hat{x}_{j},\hat{y}_{j},\hat{i}_{j}\right)\right\} \coloneqq\operatorname*{\mathtt{argmaxlocal}}_{x,y}\operatorname*{\mathtt{argmax}}_{i}\operatorname{DoG}\left(x,y,i\right)$
    \Ensure{DOF $\coloneqq \left| \left\{ \left(\hat{x}_{j},\hat{y}_{j},\hat{i}_{j}\right)\right\} \right|$}
  \end{algorithmic}
\end{algorithm}

\section{Histologic Feature Detection as a Service}\label{sec:service}

A key challenge to using our histologic feature detector is that it requires powerful GPUs with
large quantities of RAM, something that many commodity GPUs and edge devices lack.
To make our detector generally accessible we have deployed it as an on-demand service using
the funcX platform~\cite{funcx}. funcX is a high performance function-as-a-service platform designed to provide secure, fire-and-forget
remote execution. funcX federates access to remote research cyberinfrastructure via a single, multi-tenant cloud service. 
%outsourcing the
%authentication and authorization decisions to a reliable and secure Web service.
Users submit a function
invocation request to funcX which then routes the request
to the desired \emph{endpoint} for execution. Endpoints may be deployed by users on remote computing resources, including clouds, clusters, and edge devices. 

We registered our MHFD tool as a funcX function, configuring it such that it requires as input arguments only the location of the input image. The function executes the MHFD tool on an accessible GPU and the resulting feature count and DOF is returned asynchronously via the funcX service. Registration as a funcX function allows others to execute the tool on their own funcX endpoints.

We enable automated invocation of the MHFD
via Globus Flows~\cite{ananthakrishnan18platform}---a research automation platform. funcX is accessible as a Flows Action Provider, enabling users to deploy a flow that detects data creation, transfers data from instrument to analysis cluster, executes the MHFD, and returns results to users.

% Finally,a benefit of serving our histologic feature detector
% as a funcX function is that it can be leveraged within research automation pipelines.
% The funcX platform is exposed as a Globus Flows~\cite{ananthakrishnan18platform} Action Provider, meaning users can deploy an automated pipeline to move images, determine the degree of focus, and raise an alert
% when an image is deemed inappropriate.

\section{Evaluation\label{sec:Evaluation}}

% % \ryan{Not sure where this belongs. Maybe we can move to the science section?}
% Brains were prepared in the same manner and as previously described
% \cite{doi.org/10.1038/ncomms8923}. Briefly, an anesthetized animal was first transcardially
% perfused with 10ml 0.1 M Sodium Cacodylate (cacodylate) buffer, pH
% 7.4 (Electron microscopy sciences (EMS) followed by 20 ml of fixative
% containing 2\% paraformaldehyde (EMS), 2.5\% glutaraldehyde (EMS)
% in 0.1 M Sodium Cacodylate (cacodylate) buffer, pH 7.4 (EMS). The
% brain was removed and placed in fixative for at least 24 hours at
% 4C. A series of 300 um vibratome sections were prepared and put into
% fixative for 24 hours at 4C. The primary visual cortex (V1) was identified
% using areal landmarks and reference atlases. A small piece (~2 x
% 2 mm) containing V1 was cut out and prepared for EM by staining sequentially
% with 2\% osmium tetroxide (EMS) in cacodylate buffer, 2.5\% potassium
% ferrocyanide (Sigma-Aldrich), thiocarbohydrazide, unbuffered 2\% osmium
% tetroxide, 1\% uranyl acetate, and 0.66\% Aspartic acid buffered Lead
% (II) Nitrate with extensive rinses between each step with the exception
% of potassium ferrocyanide. The tissue was then dehydrated in ethanol
% and propylene oxide and infiltrated with 812 Epon resin (EMS, Mixture:
% 49\% Embed 812, 28\% DDSA, 21\% NMA, and 2.0\% DMP 30). The resin-infiltrated
% tissue was cured at 60oC for 3 days. 

We evaluate our optimized histologic feature detector in terms of runtime performance (in order to assess its fitness a realtime OOF detector).
Data used herein were collected using brains prepared in the
same manner and as previously described~\cite{doi.org/10.1038/ncomms8923}.
Using a commercial ultramicrotome
(Powertome, RMC), the cured block was trimmed to a \textasciitilde1.0mm x 1.5 mm
rectangle and \textasciitilde2,000, 40nm thick sections were collected on polyimide
tape (Kapton) using an automated tape collecting device (ATUM, RMC)
and assembled on silicon wafers as previously described~\cite{kasthuri2015saturated}. Images
at different focal distances were acquired using backscattered electron
detection with a Gemini 300 scanning electron microscope (Carl Zeiss),
equipped with ATLAS software for automated imaging. Dwell times for
all datasets were 1.0 microsecond.

\vspace{-2em}
\begin{table}
  \caption{Test platform (ALCF ThetaGPU)}
  % \vspace{-2ex}
  \centering %
  \begin{tabular}[t]{p{0.15\linewidth}p{0.75\linewidth}}
    \hline
    CPU      & Dual AMD Rome 7742 @ 2.25GHz \tabularnewline
    GPU      & 8x NVIDIA A100-40GB \tabularnewline
    HD       & 4x 3.84 U.2 NVMe SSD \tabularnewline
    RAM      & 1TB \tabularnewline
    Software & CuPy-8.3.0, CUDA-11.0, NVIDIA-450.51.05 \tabularnewline
    \hline
  \end{tabular}\label{tab:test}
\end{table}
\vspace{-2em}

We perform runtime experiments across a range of parameters of interest
(section resolution, number of feature scales). Our test platform
is a NVIDIA DGX A100 (see Table~\ref{tab:test}). Experiments consist
of computing the DOF of a sample section for a given configuration.
All experiments are repeated $k$ times (with $k=21$) and all metrics
reported are median statistics, where we discard the first execution as it is an outlier due to various
initializations (e.g,. pinning CUDA memory).

For a section resolution of $1024\times1024$ pixels we achieve approximately
a 50Hz runtime in the single GPU configuration; this is near-real-time.
We observe that, as expected, runtime grows linearly with the number
of feature scales and quadratically with the resolution of the section;
naturally, this is owing to the parallel architecture of the GPU.
The principle defect of our technique is that it is highly dependent
on the available RAM of the GPU on which it is deployed. In practice, most
GPUs available at the edge, i.e., proximal to microscopy instruments,
will have insufficient RAM to accommodate large section resolutions
and wide feature scale ranges. In fact, even the 40GB of the DGX's
A100 is exhausted at resolutions above $4096\times4096$ for more
than approximately 20 feature scales.

\begin{figure}
  \centering 
  \begin{subfigure}[t]{0.47\textwidth}
    \includegraphics[width=1\linewidth]{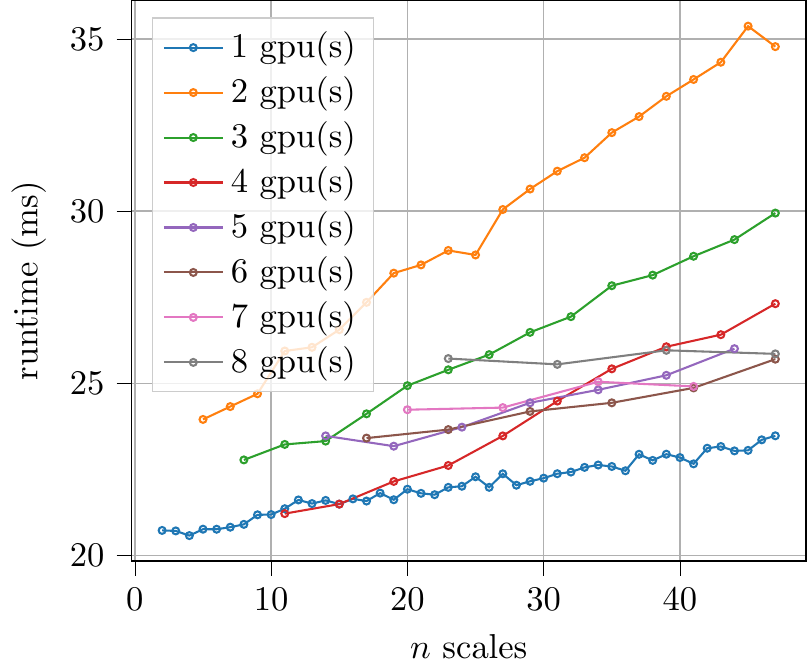}
    \caption{Median runtime as a function of number of feature scales at resolution
      $=1024\times1024$.}
    \label{subfig:nbins} 
    \end{subfigure}
  \hfill
  \begin{subfigure}[t]{0.49\textwidth}
    \includegraphics[width=1\linewidth]{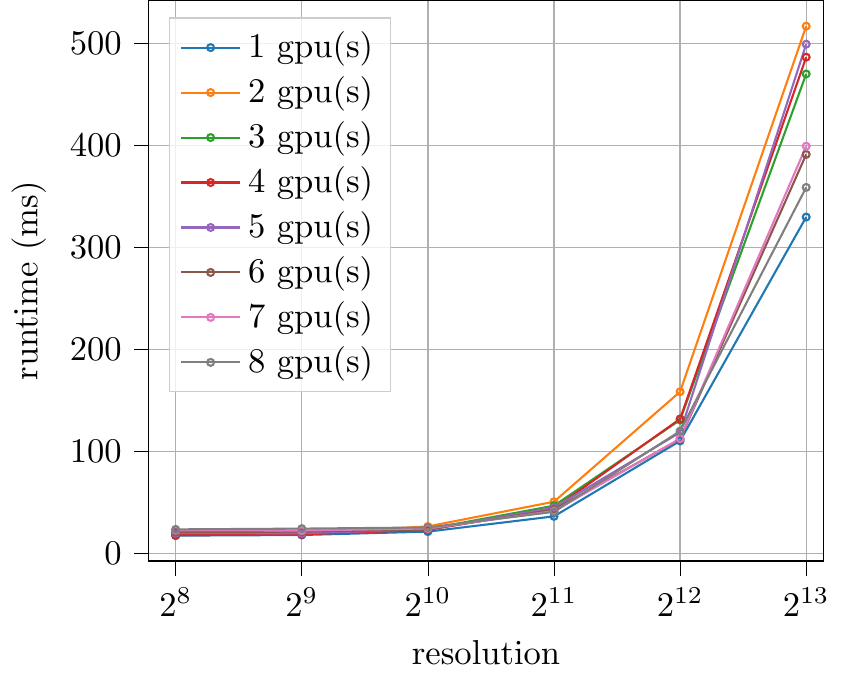}
    \caption{Median runtime as a function of section resolution with 16 feature
      scales.}
    \label{subfig:res} \end{subfigure} \caption{Scaling experiments for runtime with respect to number of GPUs, resolution,
    and number of feature scales.}
  \label{fig:evalplots}
\end{figure}

Therefore, we further investigate parallelizing MHFD across multiple
GPUs. Our implementation parallelizes MHFD in a straightforward fashion:
we partition the set of filters across the GPUs, perform the ``lighter''
FFT-IFFT pair on each constituent GPU, and then gather the results
to the root GPU (arbitrarily chosen). That is to say we actually carry
out
\[
  \left\{ L\left(x,y,t_{i}\right)\mid i\in I_{m}\right\} =\big\{\mathcal{F}^{-1}\big\{\mathcal{F}\left\{ G\left(x,y,t_{i}\right)\right\} \cdot\mathcal{F}\left\{ I\left(x,y\right)\right\} \big\}\mid i\in I_{m}\big\}
\]
where for $m=1,\dots,M$ the set $I_{m}$ indexes the scales allocated
to a node $m$. By partitioning the set of Gaussian filters $\left\{ G\left(x,y,t_{i}\right)\right\} $
across $M$ nodes, we effectively perform distributed filtering. We
use CUDA-aware OpenMPI to implement the distribution. Note that for
such multi-GPU configurations the range of feature scales was chosen
to be a multiple of the number of GPUs (hence the proportionally increasing
sparsity of data in Figure~\ref{subfig:nbins}). 

\begin{figure}
  \centering
  \includegraphics[width=0.6\linewidth]{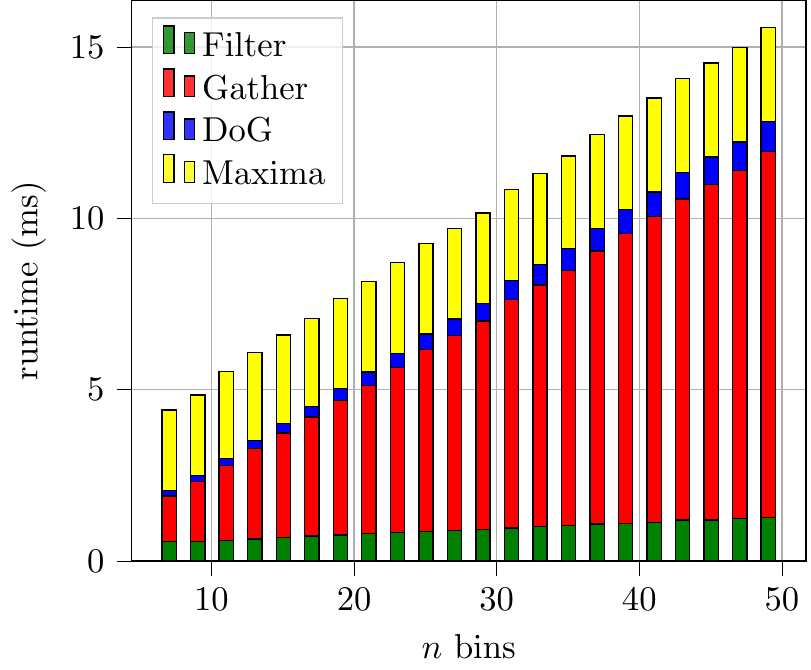}
  \caption{Breakdown of runtime into the four major phases for two GPUs across
    feature scales at resolution $=1024\times1024$.}
  \label{fig:stacked}
\end{figure}

We observe that, as one would expect, runtime is inversely proportional to number of
GPUs (see Figure~\ref{subfig:res}) but for instances where a single GPU configuration is sufficient it is also optimal. 
More precise timing reveals that parallelization across multiple GPUs incurs high network copy costs during the gather phase (see Figure~\ref{fig:stacked}).
Note that this latency persists even after taking advantage of CUDA IPC~\cite{6270863}. 
In effect, this is a fairly obvious demonstration of Amdahl's law. 
Therefore, parallelization across multiple GPUs should be considered in instances where full resolution section images are necessary (e.g., when feature scale ranges are very wide, with detection at the lower end of the scale being critical).
In all other cases, preprocessing by downsampling, by bilinear interpolation, in order to satisfy GPU RAM constraints yields a more than reasonable tradeoff between accuracy and latency.

% \section{Discussion\label{sec:Discussion}}

To evaluate the on-demand use of MHFD we deployed a funcX endpoint on the ThetaGPU cluster (Table~\ref{tab:test}).
We registered our MHFD tool as a funcX function and passed the path to the test dataset as input. 
It is important to note that the SEM imagery is not passed through the funcX platform in these tests but instead is assumed data to be resident on the shared file system.
We performed over 1000 invocations of the function after first executing a task to initialize the environment.
The tests use a single GPU and a dataset with section resolution of $1024\times1024$ pixels to be comparable to earlier results.
Our findings show that the mean time to perform the feature detection was $25$ms, with a standard deviation of $6$ms, which is similar to the previous result for the same dataset when not using funcX.
The mean time taken to submit the request to the funcX service and retrieve the result was $233$ms, with a standard deviation of $44$ms, meaning the funcX platform introduces an overhead of approximately $200$ms.
While this overhead is roughly $8\times$ greater than the MHFD analysis itself, the total time required to determine the focus of a dataset is suitable to classify images and report errors as data is collected.
In practice, data must also be moved to the ThetaGPU cluster's filesystem before it can be analyzed. 
When evaluating the time to transfer the single 23MB image from Argonne's Structural Biology Center (where the SEM resides) to the ThetaGPU cluster, we found it was moved at over 200MB/s and took roughly $100$ms. This is due to the two 40Gb/s connections between the SBC and computing facility.

% \ryan{data transfer goes here... 23MB per section tile. Moves in about 0.1s/over 200MB/s between SBC and ALCF}

%   & Time (s) & Std dev \tabularnewline
% OOF & 0.025  & 0.006 \tabularnewline
% RTT & 0.233  & 0.044 \tabularnewline

% This stuff was mentioned above and is great motivation:
% In practice, most
% GPUs available at the edge, i.e. proximal to microscopy instruments,
% will have insufficient ram to accommodate large section resolutions
% and wide feature scale ranges. In fact, even the 40GB of the DGX's
% A100 is exhausted at resolutions above $4096\times4096$ for more
% than approximately 20 feature scales.

\section{Related work\label{sec:related} }

Automating the control and optimization of scientific instruments is a common 
area of research that spans a diverse collection of devices and applies an equally diverse set of techniques, including HPC analysis, ML-in-the-loop~\cite{pan2021flame}, and edge-accelerated processing~\cite{liu2021bridging}. 
Laszewski et al.~\cite{von2000real} and Bicer et al.~\cite{bicer2017real} demonstrate two approaches to  perform real-time processing of synchrotron light source data in order to  steer experiments.
Both of these cases employ HPC systems to rapidly analyze and reconstruct data to guide instruments toward areas of interest.
Others have also used FPGAs %in addition to GPUs 
to act on streams of instrument imagery~\cite{stevanovic2015control}.
% Our work complements these as we present a novel technique to identify poor quality SEM images with the intent to alert users while data capture can be easily remedied. 

There is also much work in developing and improving auto-focus algorithms
and their applications to microscopy. 
Yeo et al.~\cite{YEO1993629} was one of the first investigations of applying auto-focus to microscopy.
They compare several criterion functions and conclude that the so-called Tenengrad criterion function is most accurate and most robust to noise.
The crucial difference between their evaluation criteria and ours is they select for criterion functions that are suited for optical microscopy, i.e., criterion functions that are robust to staining/coloring (whereas all of our samples are grayscale). Redondo et al.~\cite{10.1117/1.JBO.17.3.036008} reviews sixteen criterion functions and their computational cost in the context of automated microscopy. 
Bian et al.~\cite{https://doi.org/10.1002/jbio.202000227} address the same issues that motivate us in that they aim to support automated processes in the face of topographic variance in the samples (which leads to comparing OOF rates). 
Their solutions distinguish themselves in that they employ active devices (such as low-coherence interferometry).
Interestingly, seemingly contemporaneously with our project Luo et al.~\cite{doi:10.1021/acsphotonics.0c01774} proposed a deep learning architecture that auto-focuses in a ``single-shot'' manner. 
Such a solution is appealing given the affinity with our own application of GPGPU to the problem and we intend to experiment with applying it to our data.

\section{Conclusions}\label{sec:conclusion}
We presented an OOF detection technique designed to augment existing microscopy instrumentation.
Rather than focusing the microscope, as auto-focusing algorithms would, our algorithm operates downstream of image acquisition to report out-of-focus events to the user. 
This enables the user to intervene and initiate reacquisition protocols (on the microscope) before unknowingly proceeding with collecting the next series of images or proceeding with downstream image processing and analysis. 
Our technique is effective and operates at near-real-time latencies. 
Thus, this human-in-the-loop remediation protocol already saves the user much wasted collection time %and tedium in 
triaging defective collection runs. 
% In future work we will investigate techniques to employ our focus detection tool  in an iterative mode, in order that the DOF can be used to adjust the focus of the microscope. 
% This would require close integration with the existing software and actuation
% hardware of the microscope.

% Though our technique is efficient for use with a single GPU for small to moderately sized image sections we emphasize that distribution across multiple nodes will inevitably be necessary for use with microscopy instrumentation in the near future.
% The current state of the art ZEISS MultiSEM 505/506 employs 91 parallel electron beams and images an entire 52 tile series in approximately 1.3s~\cite{zeiss:multisem551}; this is approximately 25ms per tile or exactly 50Hz (i.e. exactly the rate at which our technique operates). 
% For maximal efficiency in the end-to-end automation of connectomics our solution (or refinements thereof) will need to be automatically employed to remove the human from the loop.

\section*{Acknowledgements}
This work was supported by the U.S. Department of Energy, Office of
Science, under contract DE-AC02-06CH11357.

\bibliographystyle{splncs04}
\bibliography{main}

\end{document}